\documentclass[preprintnumbers,showkeys,showpacs,byrevtex]{revtex4}
\usepackage{amsmath,amsfonts,amssymb,amscd,amsxtra,amsthm}
\usepackage{graphicx}
\usepackage{epstopdf}
\usepackage{bm}
\begin{document}      
\preprint{KIAS-P13020}
\title{QCD magnetic susceptibility at finite temperature beyond the chiral limit}      
\author{Seung-il Nam}
\email[E-mail: ]{sinam@kias.re.kr}
\affiliation{School of Physics, Korea Institute for Advanced Study (KIAS), Seoul 130-722, Republic of Korea}
\date{\today}
\begin{abstract}
We investigate the QCD magnetic susceptibility $\chi_q$ for flavor SU(2) at finite temperature ($T$) beyond the chiral limit, using the liquid instanton model, defined in Euclidean space and modified by the $T$-dependent caloron solution. The background electromagnetic fields are induced to the QCD vacuum, employing the Schwinger method. We first compute the scalar (chiral) and tensor condensates as functions of $T$ as well the current-quark mass $m$, signaling the correct universal chiral restoration patterns. It turns out that $\chi_q$, given by the ratio of the two condensates, is a smoothly decreasing function of $T$, showing about $20\%$ reduction of its strength at the chiral transition $T\equiv T_0$, in comparison to that at $T=0$, and decreases almost linearly beyond $T_0$ for $m\ne0$. We observe that the present numerical results are in qualitatively good agreement with other theoretical results, including the lattice simulations. Finally, we examine the effects of the external magnetic field on the tensor-polarization VEV, resulting in that it plays the role of the chiral order parameter. 
\end{abstract}
\maketitle
\section{Introduction}
The low-energy quantum chromodynamics (QCD) manifesting nonperturbative natures have been investigated extensively in many different ways, such as the lattice QCD (LQCD), effective QCD-like models, QCD sum rule (QCDSR), and so on. To understand the nonperturbative features of QCD, it is necessary to scrutinize the QCD vacuum structure, which governs the breakdown of relevant symmetries of QCD and the phase transitions of the QCD matter. Note that, recently, the electromagnetic (EM) properties of the QCD vacuum at finite temperature ($T$) and/or finite quark chemical potential ($\mu_q$) have attracted much attention from experiments~\cite{Abelev:2008ab} as well as theories~\cite{Bzdak:2011yy,Fukushima:2008xe,Tuchin:2013ie}, being together with the energetic progresses of the heavy-ion collision experiments. Among the relevant physical quantities, which are sensitive to the U(1) EM interactions, the QCD magnetic susceptibility is one of the important ones, due to its impact in theories and experiments. The QCD magnetic susceptibility for a quark flavor $q$, $\chi_q$ is stands for a response of the scalar (chiral) condensate to the external EM background field, and defined in terms of the vacuum expectation value (VEV) for the tensor-polarization operator (TP-VEV), $\langle q^{\dagger}\sigma_{\mu\nu}q\rangle_\mathrm{EM}$, in Euclidean space~\cite{Ioffe:1983ju,Kim:2004hd}:
\begin{equation}
\label{eq:VEV}
\langle q^{\dagger}\sigma_{\mu\nu}q\rangle_\mathrm{EM}
=e_q{F_{\mu\nu}}\langle{iq}^{\dagger}q\rangle \chi_q,
\end{equation}
where $e_q$ and $F_{\mu\nu}$ denote the quark electric charge and the EM field strength tensor. The subscript EM in the left-hand-side of Eq.~(\ref{eq:VEV}) stands for the existence of the external EM field. Note that TP-VEV is a linear function of $F_{\mu\nu}$ in the leading order. The magnetic susceptibility is also pertinent to the photon distribution amplitude~\cite{Ball:2002ps} as well as the anisotropy in terms of the para- or dia-magnetism~\cite{Bali:2013esa}. This physical quantity was studied up to now in QCDSR~\cite{Belyaev:1984ic,Balitsky:1985aq,Braun:2002en}, effective quark models~\cite{Kim:2004hd,Frasca:2011zn}, holographic QCD (hQCD)~\cite{Bergman:2008sg,Gorsky:2009ma},  operator product expansion with the pion dominance (OPE$+$PD)~\cite{Vainshtein:2002nv}, and LQCD~\cite{Buividovich:2009ih,Braguta:2010ej,Bali:2012jv}, and so on. Note that, in the previous work, we computed $\chi_q$ using the liquid instanton model (LIM)~\cite{Schafer:1995pz,Diakonov:2002fq} at finite density~\cite{Nam:2008ff}, in which we noticed that $\chi_q>0$ with the Euclidean metric, indicating the diamagnetic matter ($\chi_q<0$ for the Minkowski metric). Taking the present heavy-ion collision experiments at low quark density into account, in the present work, we are focusing on the QCD magnetic susceptibility for the flavor SU(2), SU($2_f$), at finite $T$ and $\mu_q=0$, using LIM, modified by the $T$-dependent caloron solution~\cite{Nam:2011vn,Nam:2013fpa}. We mention that all the calculations are performed in Euclidean space. The relevant instanton parameters, such as the average (anti)instanton size $\bar{\rho}$ and inter-(anti)instanton distance $\bar{R}$ are modified as functions of temperature, employing the trivial-holonomy (Harrington-Shepard) caloron solution~\cite{Harrington:1976dj,Diakonov:1988my,Nam:2009nn}. As a result, these modifications show the partial restorations of the spontaneous breakdown of the chiral symmetry (SB$\chi$S) as $T$ increases. Relating to those model parameters, it is worth mentioning that our model (renormalization) scale is taken as $\mu\approx\sqrt{2}/\bar{\rho}\approx0.85$ GeV at $T=0$~\cite{Nam:2011vn}. The $T$-dependent effective quark mass is computed numerically as an order parameter for SB$\chi$S, employing the $T$-modified effective thermodynamic potential~\cite{Nam:2009nn}. By doing that, we observe correct universal chiral restoration patterns for the zero and finite current-quark masses, i.e. the second-order and crossover chiral restoration patterns. Being equipped with those ingredients, we present the numerical results for the chiral condensates $\langle i{q}^\dagger q\rangle$, tensor condensate  $\langle q^\dagger\Sigma q\rangle$, their ratio corresponding to the magnetic susceptibility $\langle q^\dagger\Sigma q\rangle/\langle i{q}^\dagger q\rangle\equiv\chi_q$, and TP-VEV $\langle q^{\dagger}\sigma_{\mu\nu}q\rangle_\mathrm{EM}$. Here, we write the definition of the tensor condensate as follows:
\begin{equation}
\label{eq:MAGCON}
\langle q^\dagger\Sigma q\rangle\equiv\frac{\langle q^{\dagger}\sigma_{\mu\nu}q\rangle_\mathrm{EM}}{e_qF_{\mu\nu}}=\chi_q\langle iq^\dagger q\rangle\,\,\,\,\mathrm{for}\,\,\,\,F_{\mu\nu}\ne0.
\end{equation}
For instance, in Ref.~\cite{Bali:2012jv}, the authors defined the tensor condensate as $\langle q^\dagger\Sigma q\rangle\equiv-\tau_q$. 

From the numerical calculations, it turns out that the chiral condensate exhibits the correct chiral restoration patterns as expected from those of the effective quark mass as mentioned above. The chiral transition $T$ is given as $T_0=(166,170)$ MeV for the zero and finite current-quark masses, i.e. $m=0$, and $5$ MeV, considering the SU($2_f$) flavor symmetry. Note that the values for the chiral condensates are obtained as $\langle iq^\dagger q\rangle=(258,260\,\mathrm{MeV})^3$ for $m=(0,5)$ MeV at $T=0$. These values are well compatible with its empirical one $(240\pm10\,\mathrm{MeV})^3$. Using the same parameter sets, we also compute the tensor condensates, which also manifest the universal chiral restoration patterns. At zero temperature, we have $\langle q^\dagger\Sigma q\rangle=(52,49)$ MeV for $m=(0,5)$ MeV at $T=0$. Interestingly, we observe bump structures in the tensor condensate curves for $m\ne0$ at $T=(50\sim60)$ MeV, due to the nontrivial interference between the constituent- and current-quark masses. These values are again well-compatible with the known theoretical estimations~\cite{Kim:2004hd,Frasca:2011zn,Bali:2012jv,Belyaev:1984ic,Balitsky:1985aq,Ball:2002ps,Frasca:2011zn}. When we compare the temperature-dependent behaviors of the present numerical results with the lattice data~\cite{Bali:2012jv}, extrapolated to the physical $u$-quark mass, at the renormalization scale $\mu=1$ GeV, we observe qualitatively good agreement with them, but sizable deviations also appear in the vicinity of the chiral transition temperature $T_0$. At the chiral phase transition $T$, $T_0=170$ MeV, the tensor condensate for $m\ne0$ becomes $21$ MeV, whereas it is zero for the chiral limit. 

 It turns out that the magnetic susceptibility decrease steadily with respect to $T$. We obtain their typical values, estimated as $\chi_q=(3.03,2.77)\,\mathrm{GeV}^{-2}$ for $m=(0,5)$ MeV at $T=0$. Again, these values are well matched with those from the LQCD and other effective models. Beyond $T_0=170$ MeV, the curve for the magnetic susceptibility behave almost as a linearly decreasing one as $T$ increases. At $T_0$, we observe about $20\%$ decreases in their strengths, in comparison to those at $T=0$. TP-VEV is a linear function of the external magnetic field ($e_qB$) in the leading order. The slope of the TP-VEV line with respect to $e_qB$ decreases as $T$ increases, since the tensor condensate plays the role for its slope value, signaling the (partial) restoration of SB$\chi$S. Thus, TP-VEV can be considered as a chiral order parameter. Consequently, we observe that it vanishes at $T_0=166$ MeV in the chiral limit, whereas remains finite beyond $T_0$ for $m\ne0$, because of the crossover chiral phase transition.

We organize the present work as follows: In Section II, we briefly introduce the liquid instanton model (LIM) and how to compute the magnetic susceptibility in terms of the field theoretical manner. In Section III, the temperature modifications of the relevant model parameters are performed using the trivial caloron solution. We also show the correct universal chiral restoration patterns, computed within the present model.  The numerical results for the chiral and tensor condensates as functions of temperature are presented with relevant discussions in Section IV. In addition, the magnetic susceptibility is estimated and compared with other theoretical estimations. Final Section is devoted to summary, conclusion, and future perspectives.

\section{Effective action via the instanton-vacuum configuration}
In this Section, we introduce the liquid-instanton model (LIM) briefly as a theoretical framework to study the magnetic susceptibility. Details on the present framework can be found in Refs.~\cite{Diakonov:2002fq,Nam:2009nn}. The effective action for SU($2_f$) via the instanton vacuum can be written  in Euclidean momentum as follows~\cite{Diakonov:2002fq,Kim:2004hd}:
\begin{equation}
\label{eq:EA}
\mathcal{S}_\mathrm{eff}[m,A_\mu,T_{\mu\nu}]=-\mathrm{Sp}_{c,f,\gamma}
\ln\left[i\rlap{\,/}{D}+i\hat{m}+iM(\partial^2)+\sigma\cdot T \right],
\end{equation}
where $\mathrm{Sp}_{c,f,\gamma}$ represent the functional trace running over the color ($c$), flavor ($f$), and Lorentz index ($\gamma$). The U(1) covariant derivative reads $iD_\mu=i\partial_\mu+e_qA_\mu$, in which $e_q$ stands for the electric charge of a quark. $\hat{m}$ stands for the current-quark mass matrix, $\mathrm{diag}(m_u,m_d)$. Throughout this work, we assume the SU($2_f$) symmetry for the quark masses, i.e. $m_u\approx m_d\approx m=5$ MeV for the cases beyond the chiral limit. The effective quark mass $M(\partial^2)$ is generated from the nontrivial interactions between the quarks and (anti)instanton via the quark zero mode~\cite{Diakonov:2002fq}, and it reads
\begin{equation}
\label{eq:MDEQM}
M(\partial^2)=M_0F^2(\partial^2)=M_0\left[\frac{2}{2+\bar{\rho}^2|\partial|^2} \right]^2
\to M(k^2)= M_0\left[\frac{2}{2+\bar{\rho}^2k^2} \right]^2\equiv M_k.
\end{equation}
Here, $\bar{\rho}$ denotes the average (anti)instanton size $\sim1/3$ fm~\cite{Diakonov:2002fq}, and $M_0$ indicates the constituent-quark mass at zero virtuality. In the last step of Eq.~(\ref{eq:MDEQM}), we wrote the effective mass in the Euclidean momentum space. We defined the antisymmetric tensor $\sigma_{\mu\nu}=i(\gamma_{\mu}\gamma_{\nu}-\gamma_{\nu}\gamma_{\mu})/2$ with the external tensor source field $T^{\mu\nu}$. From the effective action in Eq.~(\ref{eq:EA}), one can write the chiral condensate space by performing the functional derivative of the effective action with respect to $m$ as follows:
\begin{equation}
\label{eq:CC}
\langle iq^\dagger q\rangle=4N_c\int_k
\left[\frac{\bar{M}^2_k}{k^2+\bar{M}^2_k}-\frac{m}{k^2+m^2} \right],
\end{equation}
where we assign as $\int_k\equiv\int\frac{d^4k}{(2\pi)^4}$ for convenience. We also have used a simplified notation $\bar{M}_k=m+M_k$. Employing the phenomenological values for the model parameters $1/\bar{\rho}\approx600$ MeV and $M_0\approx350$ MeV, which are chosen to reproduce the pion weak-decay constant $F_\pi\approx93$ MeV, we obtain $\langle iq^\dagger q\rangle\approx(250\,\mathrm{MeV})^3$ in the chiral limit. Note that this value is well compatible with its empirical values~\cite{Beringer:1900zz}. 

Similarly, the matrix element in the left-hand-side of Eq.~(\ref{eq:VEV}), i.e. TP-VEV can be evaluated by performing the functional derivative with respect to $T^{\mu\nu}$ in the presence of the EM background field, induced by the Schwinger method~\cite{Schwinger:1951nm,Kim:2004hd,Nam:2008ff}:
\begin{equation}
\label{eq:VEV1}
\underbrace{\langle q^\dagger \sigma_{\mu\nu}q\rangle_\mathrm{EM}}
_\mathrm{TP\mbox{-}VEV}
=\frac{1}{N_f}\frac{\partial\mathcal{S}_\mathrm{eff}[m,A_\mu,T^{\mu\nu}]}{\partial T^{\mu\nu}}\Big|_{T=0}.
\end{equation}
By expanding Eq.~(\ref{eq:VEV1}) in terms of $e_q$, we obtain the following expression in the leading order $\propto\mathcal{O}(e_q)$, according to $e^n_q\ll1$ for $n\ge2$~\cite{Kim:2004hd,Nam:2008ff}:
\begin{equation}
\label{eq:TME}
\langle q^{\dagger}\sigma_{\mu\nu}q\rangle_\mathrm{EM}=4N_c(e_qF_{\mu\nu})
\int\frac{d^4k}{(2\pi)^4}\left[\frac{\bar{M}_k+N_k}{(k^2+\bar{M}^2_k)^2}
-\frac{m}{(k^2+m^2)^2}\right]
\equiv e_qF_{\mu\nu}\langle q^\dagger \Sigma\, q\rangle.
\end{equation}
Here, we assign $\langle q^\dagger \Sigma\, q\rangle$ as a {\it tensor} condensate for convenience as mentioned previously. Here, $\Sigma$ stands for a scalar operator to satisfy Eq.~(\ref{eq:TME}), and its analytic form does not make any impact on the final results of the present work. It is worth mentioning that the tensor condensate is defined alternatively as $\langle q^\dagger \Sigma\, q\rangle\equiv-\tau_q$ with the Minkowski metric~\cite{Bali:2012jv}. The mass-derivative term $N_k$ is defined as  
\begin{equation}
\label{eq:TME1}
N_k\equiv-k^2\frac{\partial M_k}{\partial k^2}
=\frac{8M_0(\bar{\rho}^2k^2)}{(2+\bar{\rho}^2k^2)^3}.
\end{equation}
Notice that TP-VEV in Eq.~(\ref{eq:TME}) is a linear function of the field-strength tensor $F_{\mu\nu}$ in the leading order expansion of $e_q$. This observation is consistent with other theoretical studies~\cite{Ioffe:1983ju}. Details of the derivation of Eqs.~(\ref{eq:TME}) and (\ref{eq:TME1}) can be found in our previous work~\cite{Nam:2008ff} and references therein. Substituting Eqs.~(\ref{eq:TME}) and (\ref{eq:CC}) into Eq.~(\ref{eq:VEV}), we have the following equation for the QCD magnetic susceptibility as follows:
\begin{equation}
\label{eq:CHI}
\chi_q=\frac{\langle q^\dagger \Sigma\,q\rangle}
{\langle iq^\dagger q\rangle}=\left\{
\int_k\left[\frac{\bar{M}_k+N_k}{(k^2+\bar{M}^2_k)^2}
-\frac{m}{(k^2+m^2_q)^2}\right]\right\}\left\{\int_k
\left[\frac{\bar{M}_k}{k^2+\bar{M}^2_k}-\frac{m}{k^2+m^2_q} \right]\right\}^{-1}.
\end{equation}

\section{Temperature-dependent effective quark mass for flavor SU($2_f$)}
In this Section, we would like to briefly discuss how to compute the effective quark mass $M_0$ in Eq.~(\ref{eq:MDEQM}) as a function of $T$, and to develop the $T$-dependences for the model parameters. In Refs.~\cite{Nam:2009nn}, we derived it by using the caloron distribution with trivial holonomy, i.e. Harrington-Shepard caloron~\cite{Harrington:1976dj,Diakonov:1988my}. Firstly, we want to explain briefly how to modify $\bar{\rho}$ and $\bar{R}$ as functions of $T$, using the caloron solution. Details can be found in Ref.~\cite{Nam:2009nn}.  An instanton distribution function for arbitrary $N_c$ and $N_f$ can be written with a Gaussian suppression factor as a function of $T$ and an arbitrary instanton size $\rho$ for pure-glue QCD~\cite{Diakonov:1988my}:
\begin{equation}
\label{eq:IND}
d(\rho,T)=\underbrace{C_{N_c}\,\Lambda^b_{\mathrm{RS}}\,
\hat{\beta}^{N_c}}_\mathcal{C}\,\rho^{b-5}
\exp\left[-(A_{N_c}T^2
+\bar{\beta}\gamma n\bar{\rho}^2)\rho^2 \right].
\end{equation}
We note that the CP-invariant vacuum was taken into account in Eq.~(\ref{eq:IND}), and we assumed the same analytical form of the distribution function for both the instanton and anti-instanton. Note that the instanton number density (packing fraction) $N/V\equiv n\equiv1/\bar{R}^4$ and $\bar{\rho}$ have been taken into account as functions of $T$ implicitly. For simplicity, we take the numbers of the anti-instanton and instanton are the same, i.e. $N_I=N_{\bar{I}}=N$. We also assigned the constant factor in the right-hand-side of the above equation as $\mathcal{C}$ for simplicity. The abbreviated notations are also given as:
\begin{eqnarray}
\label{eq:PARA}
\hat{\beta}&=&-b\ln[\Lambda_\mathrm{RS}\rho_\mathrm{cut}],\,\,\,\,
\bar{\beta}=-b\ln[\Lambda_\mathrm{RS}\langle R\rangle],\,\,\,
C_{N_c}=\frac{4.60\,e^{-1.68\alpha_{\mathrm{RS}} Nc}}{\pi^2(N_c-2)!(N_c-1)!},
\cr
A_{N_c}&=&\frac{1}{3}\left[\frac{11}{6}N_c-1\right]\pi^2,\,\,\,\,
\gamma=\frac{27}{4}\left[\frac{N_c}{N^2_c-1}\right]\pi^2,\,\,\,\,
b=\frac{11N_c-2N_f}{3}.
\end{eqnarray}
Note that we defined the one-loop inverse charge $\hat{\beta}$ and $\bar{\beta}$ at certain phenomenological cutoff $\rho_\mathrm{cut}$ and $\langle R\rangle\approx\bar{R}$. $\Lambda_{\mathrm{RS}}$ denotes a scale, depending on a renormalization scheme, whereas $V_3$ for the three-dimensional volume. Using the instanton distribution function in Eq.~(\ref{eq:IND}), we can compute the average value of the instanton size $\bar{\rho}^2$ straightforwardly as follows~\cite{Schafer:1996wv}:
\begin{equation}
\label{eq:rho}
\bar{\rho}^2(T)
=\frac{\int d\rho\,\rho^2 d(\rho,T)}{\int d\rho\,d(\rho,T)}
=\frac{\left[A^2_{N_c}T^4
+4\nu\bar{\beta}\gamma n \right]^{\frac{1}{2}}
-A_{N_c}T^2}{2\bar{\beta}\gamma n},
\end{equation}
where $\nu=(b-4)/2$. It can be easily shown that Eq.~(\ref{eq:rho}) satisfies the  following asymptotic behaviors~\cite{Schafer:1996wv}:
\begin{equation}
\label{eq:asym}
\lim_{T\to0}\bar{\rho}^2(T)=\sqrt{\frac{\nu}{\bar{\beta}\gamma n}},
\,\,\,\,
\lim_{T\to\infty}\bar{\rho}^2(T)=\frac{\nu}{A_{N_c}T^2}.
\end{equation}
Here, the second relation of Eq.~(\ref{eq:asym}) indicates a correct scale-temperature behavior at high $T$, i.e., $1/\bar{\rho}\approx\Lambda\propto T$. Substituting Eq.~(\ref{eq:rho}) into Eq.~(\ref{eq:IND}), the caloron distribution function can be evaluated further:
\begin{equation}
\label{eq:dT}
d(\rho,T)=\mathcal{C}\,\rho^{b-5}
\exp\left[-\mathcal{F}(T)\rho^2 \right],\,\,\,\,
\mathcal{F}(T)=\frac{1}{2}A_{N_c}T^2+\left[\frac{1}{4}A^2_{N_c}T^4
+\nu\bar{\beta}\gamma n \right]^{\frac{1}{2}}.
\end{equation}
The instanton packing fraction $n$ can be computed self-consistently, using the following equation:
\begin{equation}
\label{eq:NOVV}
n^\frac{1}{\nu}\mathcal{F}(T)=\left[\mathcal{C}\,\Gamma(\nu) \right]^\frac{1}{\nu},
\end{equation}
where we replaced $NT/V_3\to n$, and $\Gamma(\nu)$ stands for the $\Gamma$-function with an argument $\nu$. Note that $\mathcal{C}$ and $\bar{\beta}$ can be determined easily using Eqs.~(\ref{eq:rho}) and (\ref{eq:NOVV}), incorporating the vacuum values for $n\approx(200\,\mathrm{MeV})^4$ and $\bar{\rho}\approx(600\,\mathrm{MeV})^{-1}$: $\mathcal{C}\approx9.81\times10^{-4}$ and $\bar{\beta}\approx9.19$. Finally, in order for estimating the $T$-dependence of $M_0$, one needs to consider the normalized distribution function, defined as follows:
\begin{equation}
\label{eq:NID}
d_N(\rho,T)=\frac{d(\rho,T)}{\int d\rho\,d(\rho,T)}
=\frac{\rho^{b-5}\mathcal{F}^\nu(T)
\exp\left[-\mathcal{F}(T)\rho^2 \right]}{\Gamma(\nu)}.
\end{equation}
Here, the subscript $N$ denotes the normalized distribution. For brevity, we want to employ the large-$N_c$ limit to simplify the expression for $d_N(\rho,T)$. In this limit, as understood from Eq.~(\ref{eq:NID}), $d_N(\rho,T)$ can be approximated as a $\delta$-function: 
\begin{equation}
\label{eq:NID2}
\lim_{N_c\to\infty}d_N(\rho,T)=\delta[{\rho-\bar{\rho}(T)}].
\end{equation}

\begin{figure}[t]
\begin{tabular}{cc}
\includegraphics[width=8.5cm]{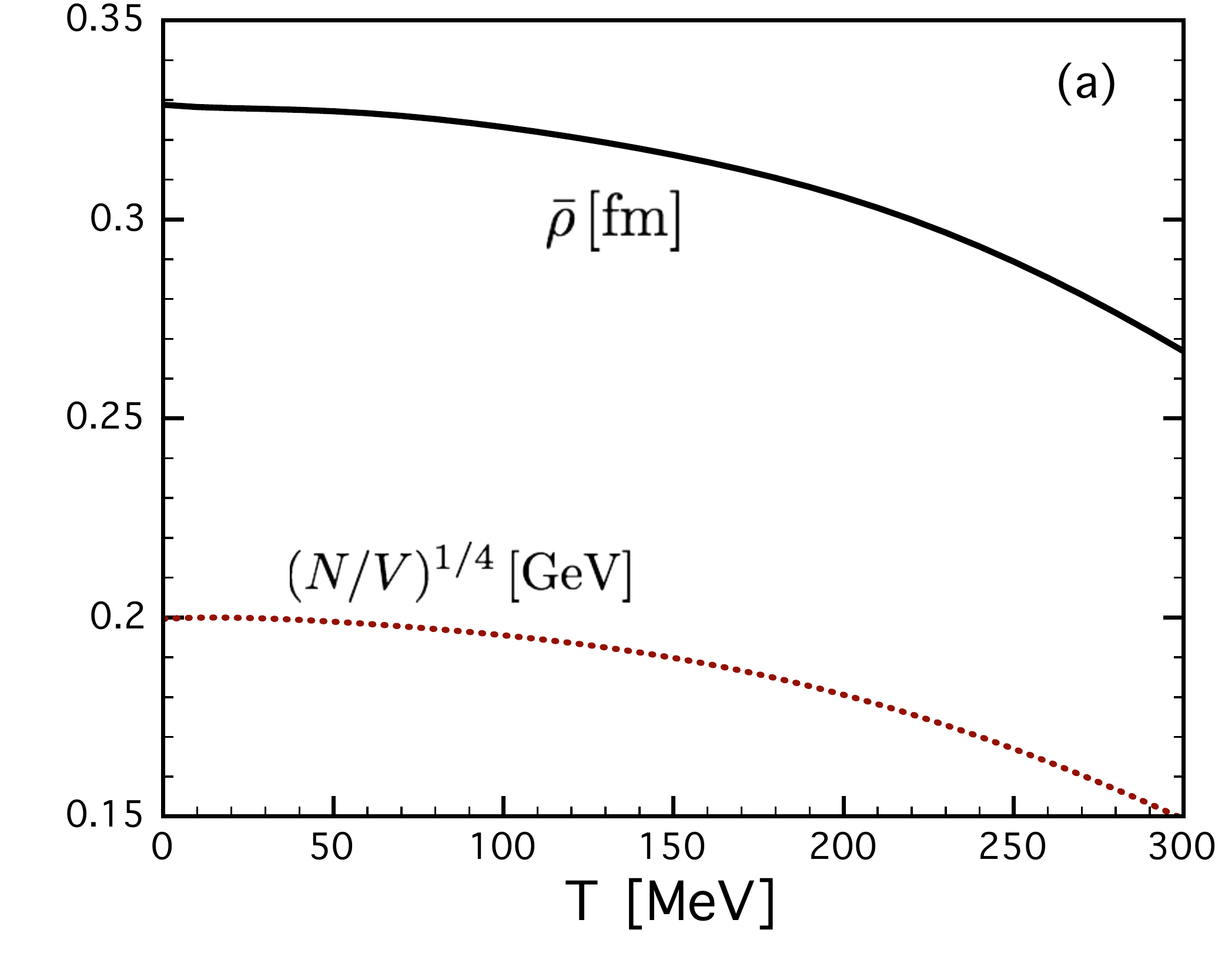}
\includegraphics[width=8.5cm]{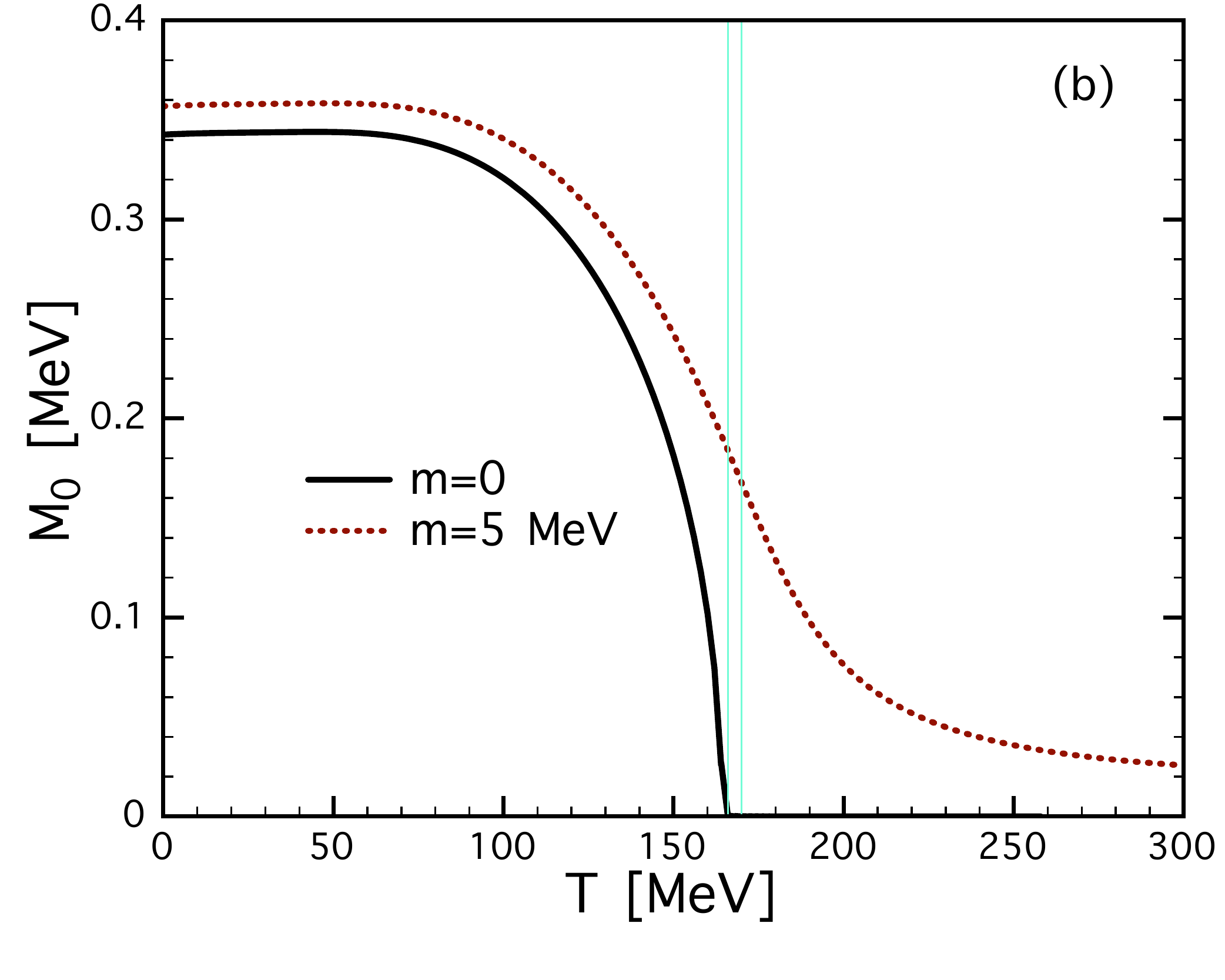}
\end{tabular}
\caption{Average (anti)instanton size $\bar{\rho}\approx1/\Lambda$ [fm] and (anti)instanton packing fraction $(N/V)^{1/4}$ [GeV] as functions of $T$, computed from the Harrington-Shepard caloron distribution~\cite{Harrington:1976dj,Diakonov:1988my} in the panel (a). Effective quark mass at zero virtuality, $M_0$ computed from Eq.~(\ref{eq:LIMGAP}) as functions of $T$ for $m=0$ (solid) and $m=5$ MeV (dot), signaling the second-order and crossover chiral phase transitions, respectively, in the panel (b). The vertical lines indicate the chiral-phase-transition temperatures $T_0=(166,170)$ MeV for $m=(0,5)$ MeV.}       
\label{FIG01}
\end{figure}

The numerical result for the trajectories of $\bar{\rho}(T)$ (solid) and $1/\bar{R}(T)$ (dot) are given in the panel (a) in Figure~\ref{FIG01}. Here we choose $\bar{\rho}(0)\approx1/3$ fm and $\bar{R}\approx1$ fm for all the numerical calculations. These values are phenomenologically preferred in the present model~\cite{Diakonov:2002fq}. The curve for $\bar{\rho}(T)$ shows that the average (anti)instanton size smoothly decreases with respect to $T$, indicating that the instanton ensemble gets diluted and the nonperturbative effects via the quark-instanton interactions are diminished. At $T=(150\sim200)$ MeV, which is close to the chiral phase transition $T$, the instanton size decreases by about $(10\sim20)\%$ in comparison to its vacuum value. Considering that the instanton size corresponds to the scale parameter of the model, i.e. UV cutoff mass, $\bar{\rho}\approx1/(\sqrt{2}\mu)$, the $T$-dependent cutoff mass is a clearly distinctive feature in comparison to other low-energy effective models, such as the NJL model. In addition, we also show the $T$ dependence of the average (anti)instanton number density or (anti)instanton packing fraction, $N/V\approx1/\bar{R}^4$, in the panel (a) of Figure~\ref{FIG01}. Again, the instanton number density get decreased as temperature increases: The instanton ensemble gets diluted. We will use these two temperature-dependent quantities for computing the chiral and tensor condensates, and TP-VEV in the next Section. 

Now, we are in a position to discuss the $T$ dependence of the constituent quark mass $M_0$ as a chiral order parameter. As in Ref.~\cite{Nam:2009nn}, the LIM thermodynamic potential per volume in the leading large-$N_c$ contributions at zero quark chemical potential can be written as follows:
\begin{eqnarray}
\label{eq:TP}
\Omega_\mathrm{LIM}
&=&\frac{N}{V}\left[1-\ln\frac{N}{\lambda V\mathrm{M}} \right]+2\sigma^2-2N_cN_f\int^\infty_0\frac{d^3\bm{k}}{(2\pi)^3}
\left[E_{\bm{k}}+2T\ln\left[1+e^{-\frac{E_{\bm{k}}}{T}}  \right]
 \right],
\end{eqnarray}
where $\lambda$ and $\mathrm{M}$ represent a Lagrange multiplier to exponentiate the effective quark-instanton action and an arbitrary massive parameter to make the argument for the logarithm dimensionless. $\sigma$ stands for the VEV for the isosinglet scalar meson field corresponding to the effective quark mass. The quark energy is defined by $E^2_{\bm{k}}=\bm{k}^2+\bar{M}^2_{\bm{k}}$. In the leading large-$N_c$ contributions, we have the relation $2\sigma^2=N/V$~\cite{Nam:2009nn}. The gap  equation can be derived from Eq.~(\ref{eq:TP}) by differentiating $\Omega_\mathrm{LIM}$ by the Lagrange multiplier $\lambda$:
\begin{equation}
\label{eq:LIMGAP}
\frac{\partial\Omega_\mathrm{LIM}}{\partial \lambda}=0\to
\frac{N_f}{\bar{M}_0}\frac{N}{V}-2N_cN_f\int^\infty_0\frac{d^3\bm{k}}{(2\pi)^3}
F^4_{\bm{k}}\frac{M_0}{E_{\bm{k}}}\left[1-\frac{2e^{-\frac{E_{\bm{k}}}{T}}}{1+e^{-\frac{E_{\bm{k}}}{T}}}\right]=0.
\end{equation}
Note that one can write the instanton packing fraction in terms of the effective quark mass $M_0$ and $\bar{\rho}$~\cite{Diakonov:2002fq}:
\begin{equation}
\label{eq:NOV}
\frac{N}{V}=\frac{\mathcal{C}_0N_cM^2_0}{\pi^2\bar{\rho}^2}.
\end{equation}
The value of $\mathcal{C}_0$ locates in $(1/3\sim1/4)$ for $1/\bar{\rho}\approx600$ MeV, $M_0\approx(300\sim400)$ MeV and $N/V\approx(200\sim260\,\mathrm{MeV})^4$ for vacuum~\cite{Goeke:2007bj}. We choose $\mathcal{C}_0=0.27$ to reproduce $M_0=(340\sim350)$ MeV at $(T,\mu)=0$ in the chiral limit. After solving Eq.~(\ref{eq:LIMGAP}) with respect to $M_0$ numerically, the numerical results for $M_0$ as a function of $T$ are given in the panel (b) of Figure~\ref{FIG01} for the zero and finite current quark mass: $m=0$ (solid) and $m=5$ MeV (dot). These results indicate correct universal patterns for the chiral phase transition like the those of the Ising model, i.e. the second-order chiral phase transition for the massless fermion and the crossover for the finite mass. From those numerical results, the phase transition $T$ for the two chiral restorations are obtained as $T_0\approx(166,170)$ MeV for $m=(0,5)$ MeV. The transition temperatures are indicated by the thin solid vertical lines in the panel (b) of Figure~\ref{FIG01}. Detailed discussions for the chiral phase structure within the present model is given in Ref.~\cite{sinam}.

To evaluate $\chi_q$ as a function of $T$, we redefine Eq.~(\ref{eq:CHI}) with the fermionic Matsubara formula: The integral over the fourth momentum, $k_4$, is compactified into a summation over the Matsubara frequency:
\begin{equation}
\label{eq:CHIT}
\chi_q=\frac{\langle q^\dagger \Sigma\, q\rangle}
{\langle iq^\dagger q\rangle}=\left\{
T\sum_{n=-\infty}^\infty\int_{\bm{k}}\left[\frac{\bar{M}_{\bm{k}}+N_{\bm{k}}}{(w^2_n+E^2_{\bm{k}})^2}
-\frac{m}{(w^2_n+E^2_0)^2}\right]\right\}\left\{
T\sum_{n=-\infty}^\infty\int_{\bm{k}}
\left[\frac{\bar{M}^2_{\bm{k}}}{w^2_n+E^2_{\bm{k}}}-\frac{m}{w^2_n+E^2_0} \right]\right\}^{-1},
\end{equation}
where the fermionic Matsubara frequency $w_n=(2n+1)\pi T$ and the three-dimensional integral is given in a simplified notation $\int_{\bm{k}}\equiv\int\frac{d^3\bm{k}}{(2\pi)^3}$. Here, we use the notation $E^2_0=\bm{k}^2+m^2$, whereas $\bm{k}$ denotes the three momentum of the quark. Here is one caveat: Introducing the Matsubara frequency, we assume that the effective quark mass in Eq.~(\ref{eq:MDEQM}) is simplified by $k_4\to0$:
\begin{equation}
\label{eq:}
M_k\to M_{\bm{k}}=M_0\left[\frac{2}{2+\bar{\rho}^2\bm{k}^2} \right]^2,
\end{equation}
and the same for $N_k\to N_{\bm{k}}$ in Eq.~(\ref{eq:TME1}). We have verified that this simplification makes the problem in hand much convenient and simplified for the analytic as well as the numerical calculations, and does not make significant deviations from the full calculations, as shown in many successful applications~\cite{Nam:2009nn,Nam:2011vn,Nam:2013fpa}. Then, the summations over $w_n$ in Eq.~(\ref{eq:CHIT}) can be analytically performed, and we defined the following functions:
\begin{equation}
\label{eq:SUM}
f_2(E)\equiv
\frac{1}{8TE^3}\mathrm{sech}^2\left(\frac{E}{2T} \right)
\left[T\mathrm{sinh}\left(\frac{E}{T} \right)-E\right],\,\,\,\,
f_1(E)\equiv
\frac{1}{4E}\mathrm{tanh}\left(\frac{E}{2T} \right).
\end{equation}
Using Eq.~(\ref{eq:SUM}), Eq.~(\ref{eq:CHIT}) can be then rewritten finally as
\begin{equation}
\label{eq:CHIT1}
\chi_q=\frac{\langle q^\dagger \Sigma\, q\rangle}
{\langle iq^\dagger q\rangle}=\left\{
\int_{\bm{k}}\left[(\bar{M}_{\bm{k}}+N_{\bm{k}})f_2(E_{\bm{k}})
-{m}f_2(E_0) \right]\right\}\left\{\int_{\bm{k}}
\left[{\bar{M}_{\bm{k}}}f_1(E_{\bm{k}})-{m}f_1(E_0)\right]\right\}^{-1}.
\end{equation}
Note that all the quantities, $\chi_q$, and chiral and tensor condensates, are positive-real  valued functions of $T$ and $m$ in the present work with the Euclidean metric. Hence, our theoretical results exhibit the diamagnetism, considering that $\chi^\mathrm{M}_q<0$ for the Minkowski (M) metric.
\section{Numerical results and discussions}
\begin{figure}[t]
\begin{tabular}{cc}
\includegraphics[width=8.5cm]{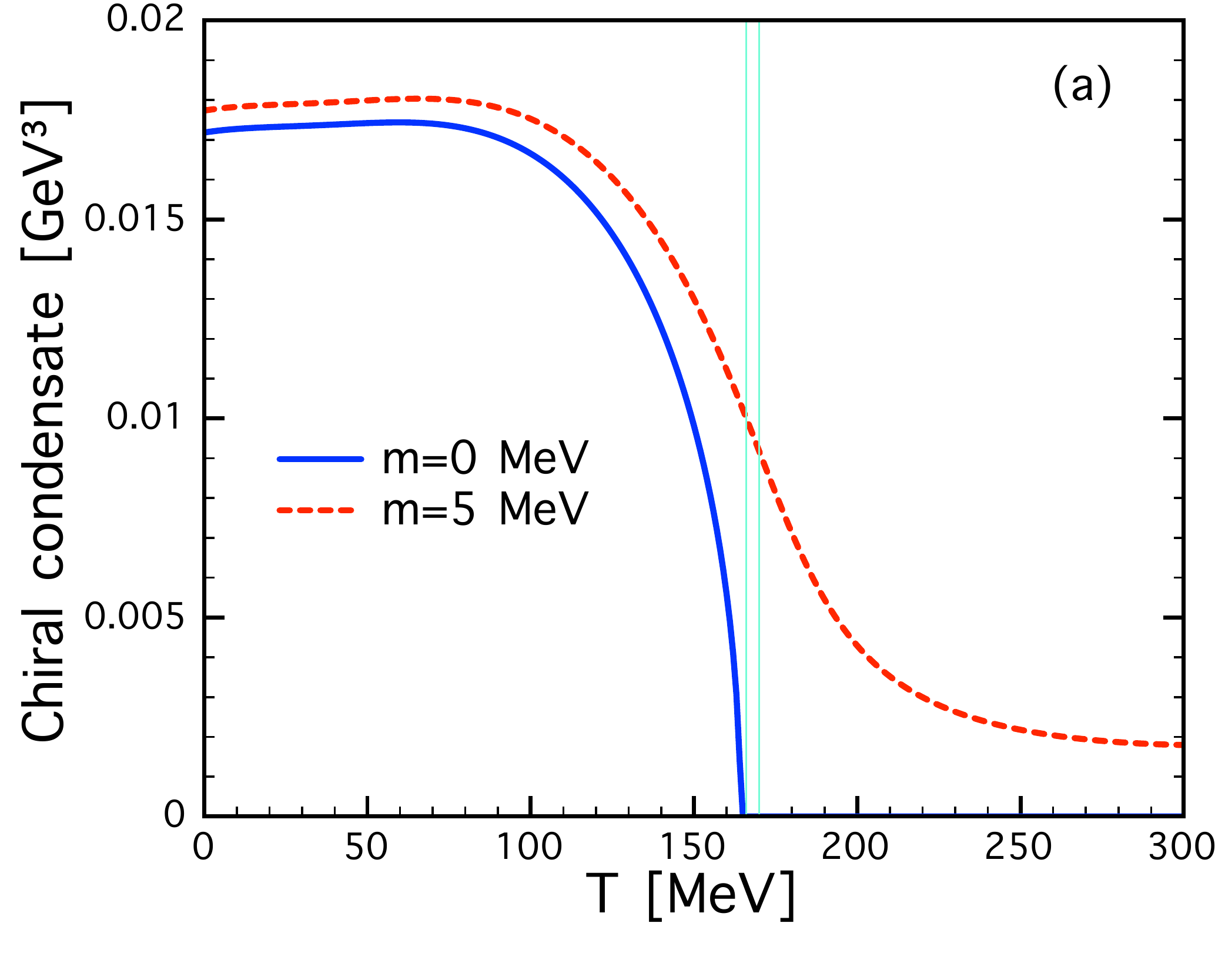}
\includegraphics[width=8.5cm]{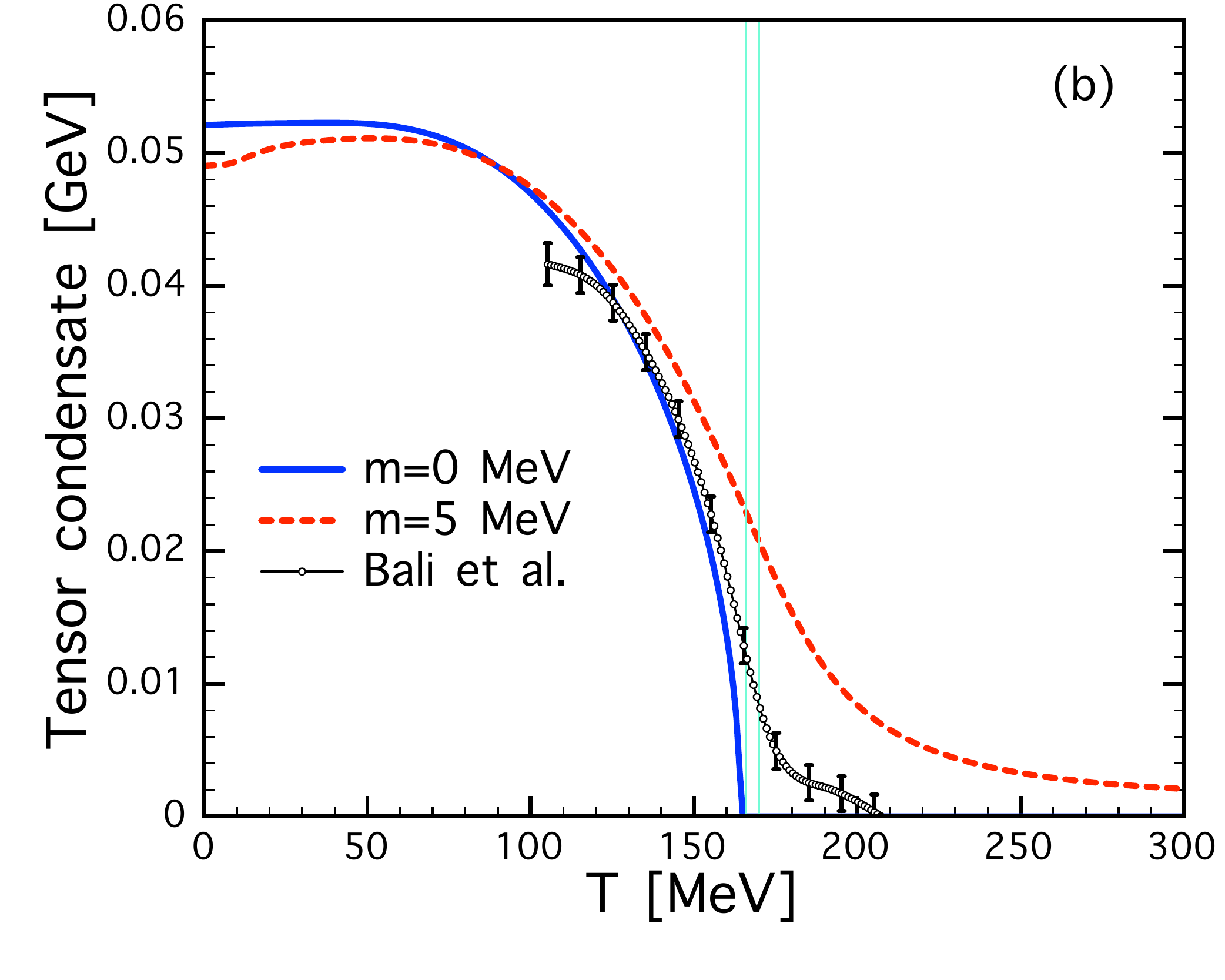}
\end{tabular}
\caption{(Color online) (a) Chiral condensate $\langle iq^\dagger q\rangle=-\langle\bar{q}q\rangle$ in Eq.~(\ref{eq:CHIT1}) for $m=(0,5)$ MeV in the (solid, dash) lines, respectively. The vertical lines indicate the chiral phase transition $T$, $T_0=(166,170)$ MeV for $(m=0,m\ne0)$. The horizontal shaded area denotes the range of $\langle iq^\dagger q\rangle=(250\sim260\,\mathrm{MeV})^3$, which corresponds to its empirical value. (b) Tensor condensate $\langle q^\dagger\Sigma q\rangle\equiv-\tau_q$ in Eq.~(\ref{eq:CHIT1}), represented in the same manner with the panel (a). The lattice QCD data at the renormalization scale $\mu=1$ GeV are taken from Ref.~\cite{Bali:2012jv}, and indicates the continuum extrapolation for the $u$ quark with the errors
containing all statistical and systematic errors. The data for the $d$ quark are still within the $u$-quark errors.}       
\label{FIG23}
\end{figure}

In this Section, we demonstrate the numerical results with relevant discussions. First, we show them for the chiral condensates $\langle iq^\dagger q\rangle$ as functions of $T$ in the panel (a) of Figure~\ref{FIG23}. The (solid, dash) curves correspond to the condensates with $m=(0,5)$ MeV. In what follows, numerical values given in the form of $(x,y)$ represents the theoretical results for $m=(0,5)$ MeV, respectively, unless otherwise stated. The vertical straight lines denote the chiral transition $T$, $T_0=(166,170)$ MeV. At $T=0$, we observe $\langle iq^\dagger q\rangle^{1/3}=(258\,\mathrm{MeV},260\,\mathrm{MeV})$ at $T=0$. Here, we only observe small deviations depending on $m$. These numerical values are slightly larger than the empirical values $\langle iq^\dagger q\rangle=(240\pm10\,\mathrm{MeV})^3$, but still in qualitative agreement with them. As for the curve for $m=0$, it shows the second-order chiral phase transition as expected and understood by Eq.~(\ref{eq:CC}), which is proportional to $M_0$. On the contrary, $m$ becomes finite, the curves manifest the crossover transition, satisfying the universal class pattern of the chiral restoration. At $T_0=170$ MeV, we have  $\langle iq^\dagger q\rangle\approx(209\,\mathrm{MeV})^3$ for $m=5$ MeV, showing about $20\%$ reduction, in comparison to that for  $T=0$.

In the panel (b) of Figure~\ref{FIG23}, we depict the tensor condensate $\langle q^\dagger \Sigma q\rangle\equiv-\tau_q$ with the same manner of the panel (a). Again, the condensates show the proper chiral restoration patterns depending on $m$. Interestingly, there appear bump structures in the curves for $m\ne0$ at $T=(50\sim60)$ MeV, due to the nontrivial interference between the constituent- and current-quark mass terms in Eq.~(\ref{eq:TME}). We find $\langle q^\dagger \Sigma q\rangle=(52,49)$ MeV at $T=0$. From these values, we conclude that the tensor condensate decreases with respect to $m$ which is consistent with the observation of Ref.~\cite{Goeke:2007nc}. Here, we want to mention other theoretical estimations for the tensor condensate for SU($2_f$) for $T=0$. Using QCDSR techniques, it was studied in Refs.~\cite{Belyaev:1984ic,Balitsky:1985aq,Ball:2002ps}, which estimated it as $(40\sim70)$ MeV, depending on the different renormalization scales $\mu=0.5$ GeV or $\mu=1$ GeV. In Ref.~\cite{Kim:2004hd}, employing the same instanton model for vacuum, with slightly different model parameters, the authors calculated it in the chiral limit, resulting in $(45\sim50)$ MeV, which is well compatible with ours by construction. Employing the NJL model and quark model (QM), it was estimated as $69$MeV and $65$ MeV, respectively, in Ref.~\cite{Frasca:2011zn} which is about $10\%$ larger than ours. From the quenched LQCD simulations for SU($2_c$)~\cite{Buividovich:2009ih} and SU($3_c$)~\cite{Braguta:2010ej}, they observed $46$ MeV and $\sim52$ MeV at $\mu=2$ GeV. A full SU($3_c$) lattice simulation was performed in Ref.~\cite{Bali:2012jv}, and provided $(38.9\sim40.7)$ for the chiral limit and physical current-quark mass at $\mu=2$ GeV. Considering that the renormalization constant for the tensor condensate with the running scale $\mu=(1\sim2)$ GeV is close to unity as will be shown below, our present estimations for the tensor condensates are well compatible with other theoretical results. The comparisons with other studies are summarized in Table~\ref{TABLE0}

Now, we are in a position to discuss the $T$ dependence of the tensor condensate. Bali {\it et al.} estimated the $T$ dependence of the tensor condensate via the full SU($3_c$) lattice simulation with tree-level Symanzik improved gauge action~\cite{Bali:2012jv}. They also performed renormalization-group analyses, resulting in the renormalization constants for the chiral (scalar) (S) and tensor (T) condensates for the running scale $\mu=(2\to1)$ GeV as follows, considering $\chi_q=\langle q^\dagger \Sigma q\rangle/\langle iq^\dagger q\rangle$:
\begin{equation}
\label{eq:RENO}
\mathrm{Z}_{\overline{\mathrm{MS}}}^\mathrm{S}\approx0.76,\,\,\,\,
\mathrm{Z}_{\overline{\mathrm{MS}}}^\mathrm{T}\approx1.13.
\end{equation}
Note that in their lattice simulation, they renormalized all the quantities at $\mu=2$ GeV with $(m_u+m_d)\approx6.94$ MeV, which gives $\langle \bar{q}q\rangle=-(269\,\mathrm{MeV})^3$. Using $\mathrm{Z}_{\overline{\mathrm{MS}}}^\mathrm{S}$ in Eq.~(\ref{eq:RENO}), this chiral condensate value becomes $-(245\,\mathrm{MeV})^3$ at $\mu=1$ GeV. Since our renormalization scale is about $0.85$ GeV $\sim\sqrt{2}/\bar{\rho}$, for appropriate comparison, one needs to evolve their data at $\mu=2$ GeV to those at $\mu=1$ GeV. After the scale evolution, just multiplying $\mathrm{Z}_{\overline{\mathrm{MS}}}^\mathrm{T}$ to the data, the continuum extrapolated data for the $u$ quark are given with the open dot with the sum of the statistical and systematic errors  in the panel (b) of Figure~\ref{FIG23}. Note that the lattice data locates between the curves for $m=0$ and $m=5$ MeV. Although there appear quantitative differences between the lattice data and ours, the overall tendency and strength are qualitatively comparable. Note that the chiral transition $T$ was given $T_0\approx162$ MeV in the lattice simulation which is a only few percent smaller than ours $T_0=170$ MeV. At the chiral phase transition $T$, $T_0=170$ MeV, the tensor condensate for $m=5$ MeV becomes $21$ MeV from our calculations, showing about two-times reduction.  

\begin{figure}[t]
\begin{tabular}{cc}
\includegraphics[width=8.5cm]{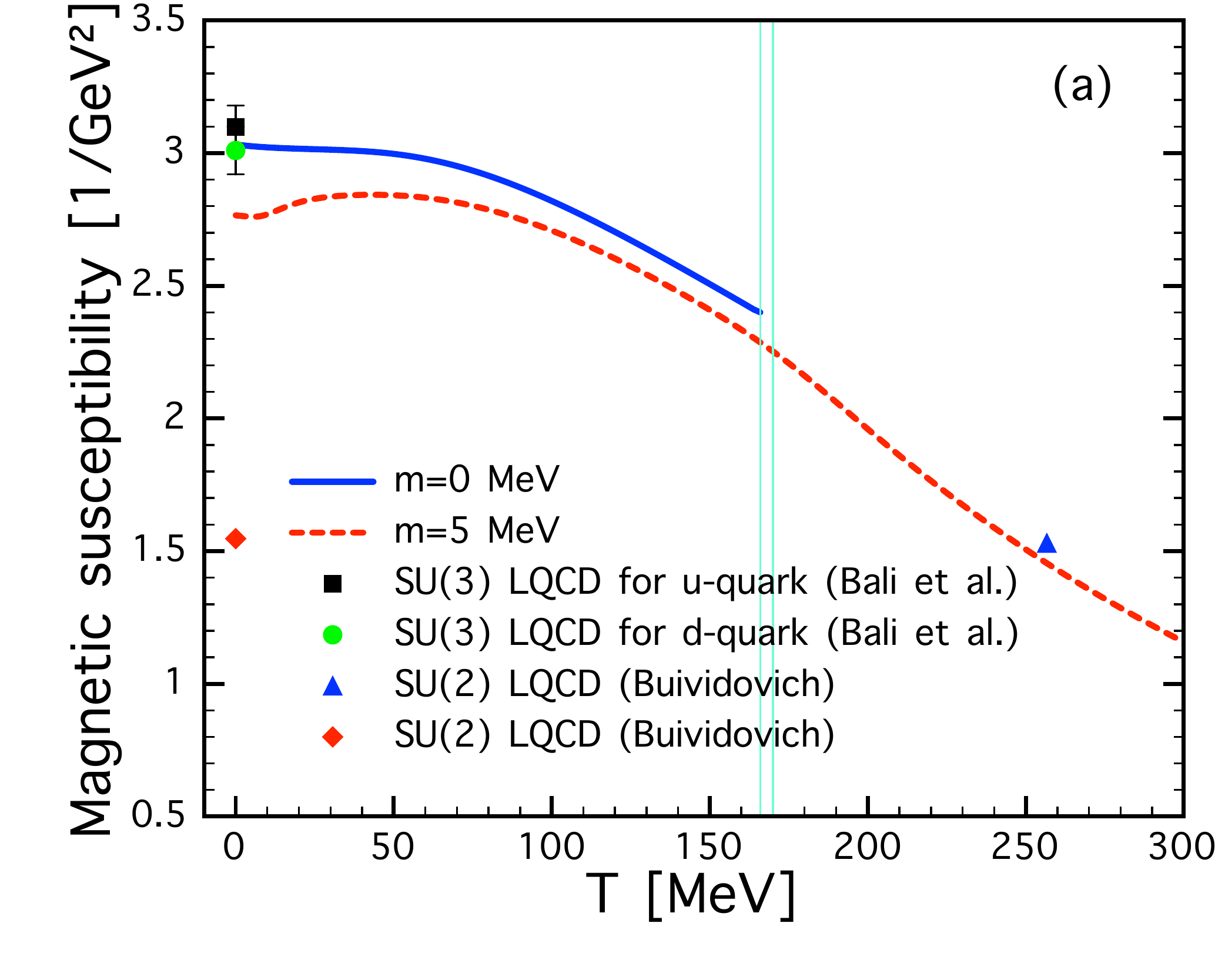}
\includegraphics[width=8.5cm]{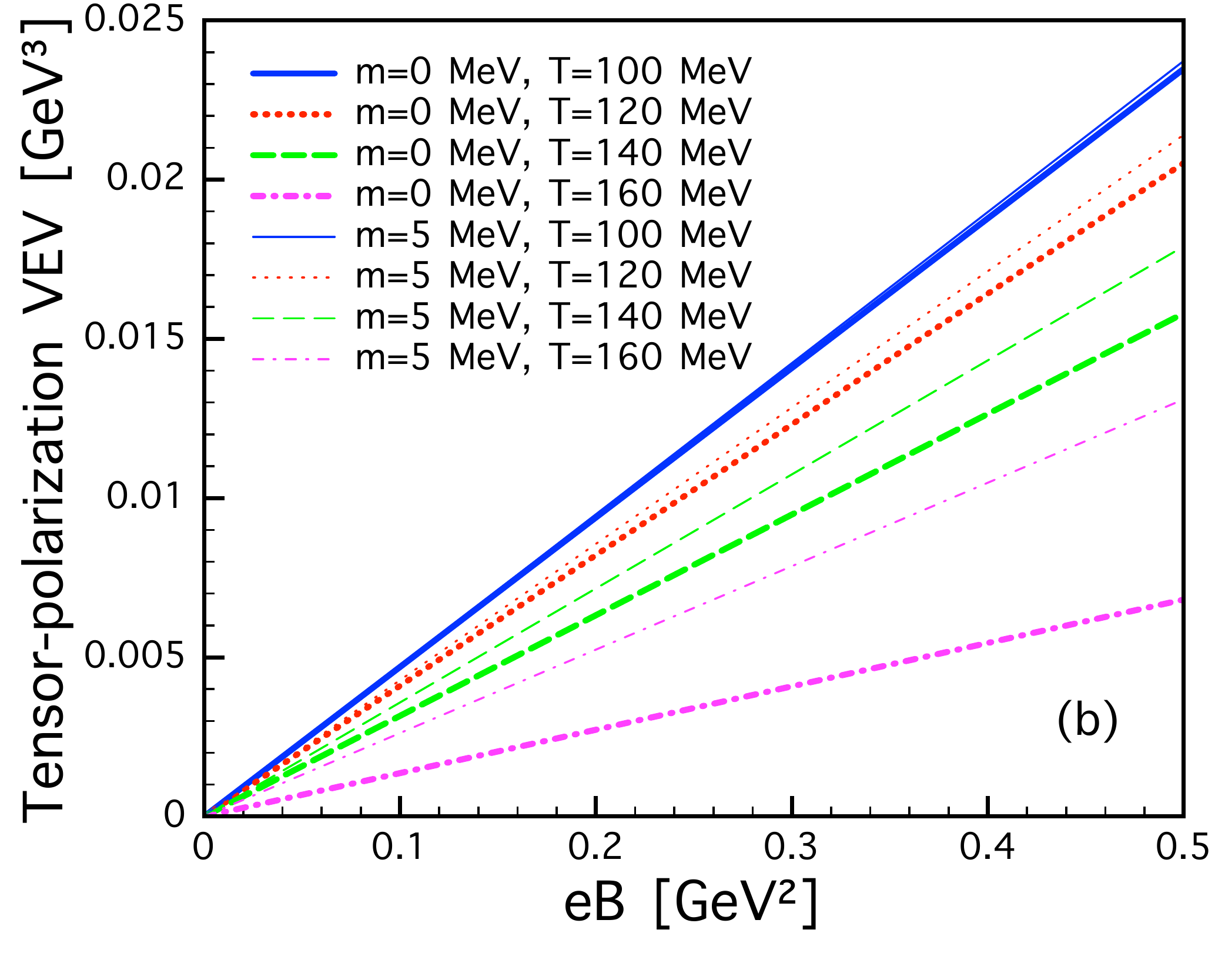}
\end{tabular}
\caption{(Color online) (a) Magnetic susceptibility $\chi_q$ as functions of $T$ for $m=(0,5)$ MeV in the (solid, dash) lines, respectively. The SU($3_c$) lattice QCD data at the renormalization scale $\mu=1$ GeV are taken from Ref.~\cite{Bali:2012jv} for the $u$-quark and $d$-quark, given in the solid square and circle. The SU($2_c$) quenched LQCD data at $\mu=2$ GeV are also shown with the solid triangle and diamond~\cite{Buividovich:2009ih}. The vertical lines indicate the chiral phase transition $T$, $T_0=(166,170)$ MeV for $m=(0,5)$ MeV. (b) Tensor-polarization VEV (TP-VEV), $\langle q^\dagger\sigma_{xy}q\rangle_\mathrm{EM}$ as functions of $e_qB$ for different $T$ and $m$ values. See the text for details.}       
\label{FIG45}
\end{figure}

In the panel (a) of Figure~\ref{FIG45}, we show the numerical results for $\chi_q$ as functions of $T$ for $m=(0,5)$ MeV in the (solid, dash) lines. Since $\chi_q$ is the ratio of the chiral and tensor condensates as shown in Eq.~(\ref{eq:CHI}), we can not define it in the chiral limit beyond $T_0$, while $\chi_q$ for $m\ne0$ has finite values for $T\ge T_0$ as shown in the panel (a). The typical values for $\chi_q$ at $T=0$ are given by $\chi_q=(3.03,2.77)\,\mathrm{GeV}^{-2}$.  Note that we have $\chi_q=(2.85\sim5.7)\,\mathrm{GeV}^{-2}$ from the QCDSR methods~\cite{Belyaev:1984ic,Balitsky:1985aq,Ball:2002ps}, and these values are well compatible with our estimations.  
The LQCD simulations also estimated comparable values with ours~\cite{Bali:2012jv,Buividovich:2009ih,Braguta:2010ej} as shown in Table~\ref{TABLE0}. However, the hQCD~\cite{Bergman:2008sg}  and OPE+PD~\cite{Vainshtein:2002nv} calculations showed considerably larger values for them: $\chi_q=11.5$ and $8.91$, respectively. The effective quark models, such as NJL and QM, evaluated $\chi_q\approx4.3$ and $5.25$, depending on each models. From these observations, we can conclude that our model estimations are compatible qualitatively with other theoretical ones. At $T=0$, we depict some LQCD results in the panel (a) of Figure~\ref{FIG45}. The solid square and circle are the estimations at $\mu=1$ GeV from the full SU($3_c$) LQCD simulation for $m=5$ after a proper scale evolution by multiplying $\mathrm{Z}_{\overline{\mathrm{MS}}}^\mathrm{T}/\mathrm{Z}_{\overline{\mathrm{MS}}}^\mathrm{S}=1.49$. It turns out that the LQCD data match with the present numerical curve for $m=0$ approximately, and larger than that for $m=5$ MeV by about $10\%$. The SU($2_c$) quenched LQCD data at $\mu=2$ GeV are also shown with the solid triangle and diamond for different $T$~\cite{Buividovich:2009ih}. Although their estimation at $0.82\,T_c$ with $T_c=313$ MeV (triangle) is comparable to ours, one must be careful about that the simulations were done for $N_c=2$ at a relatively larger renormalization scale without dynamic quarks. Note that their value at $T=0$ (diamond) is much smaller than ours as well as the SU($3_c$) LQCD simulation~\cite{Bali:2012jv}. It was suggested that the magnetic susceptibility can be parameterized in terms of OPE+PD, using the Gell-Mann--Oakes--Renner (GMOR) relation, as follows~\cite{Vainshtein:2002nv}:
\begin{equation}
\label{eq:CQ}
\chi_q=\frac{c_{\chi_q}N_c}{8\pi^2f^2_\pi}
=2.22\,c_{\chi_q}\,\mathrm{GeV}^{-2}\,\,\mathrm{for}\,\,N_c=3,
\end{equation}
where we have chosen the normalization $f_\pi=\sqrt{2}F_\pi$ with $F_\pi=92.4$ MeV and $c_{\chi_q}$ stands for a positive real constant. We list the (average) values for $c_{\chi_q}$ for the various theory calculations in Table~\ref{TABLE0}. Approximately, its value amounts to $(1.0\sim2.0)$ for various calculations, whereas the OPE+PD and hQCD still give larger values than others as the magnetic susceptibility. 
\begin{table}[b]
\begin{tabular}{c||c|c|c|c|c|c|c|c|c|c}
&Present&LIM~\cite{Kim:2004hd}&NJL~\cite{Frasca:2011zn}&LQCD$^f_3$~\cite{Bali:2012jv}&LQCD$^q_2$~\cite{Buividovich:2009ih}&LQCD$^q_3$~\cite{Braguta:2010ej}&SR~\cite{Belyaev:1984ic,Balitsky:1985aq,Ball:2002ps}&OPEPD~\cite{Vainshtein:2002nv}&hQCD~\cite{Bergman:2008sg}&QM~\cite{Frasca:2011zn}\\
\hline
$\mu$&$0.85$&$0.85$&$0.627$&$2.0$&$2.0$&$2.0$&$0$&$0.5$
&$\ll1.15$&$0.560$\\
\hline
$m$&$(0,5)$&$5$&$5$&$0,3.47$&$0$&$0$&Physical&Physical
&$0$&$5$\\
\hline
$\langle q^\dagger\Sigma q\rangle$&$(52,49)$&$40\sim45$&$69$&$38.9\sim40.7$&$46$&$\sim52$&$40\sim70$&$-$&$-$&$65$\\
\hline
$\chi_q$&$(3.03,2.77)$&$2.5\pm0.15$&$4.3$&$1.93\sim2.16$&$1.547$&$4.24\pm0.18$&$2.85\sim5.7$&$8.91$&$11.5$&$5.25$\\
\hline
$\bar{c}_{\chi_q}$&$1.26$&$1.12$&$1.94$&$0.92$&$1.045$&$1.91$&$1.95$&$4.01$&$5.18$&$2.36$\\
\end{tabular}
\caption{Various theoretical estimations for the tensor condensate $\langle q^\dagger \Sigma q\rangle$ [MeV] and magnetic susceptibility $\chi_q$ [GeV$^{-2}$] for certain renormalization scale $\mu$ [GeV] and the current-quark mass [MeV] at $T=0$. All the listed values are converted to those for Euclidean space. The notation LQCD$^{(f,q)}_{(2,3)}$ indicates the (full, quenched) LQCD simulations for $N_c=(2,3)$. $\bar{c}_{\chi_q}$ denotes the average value over possible $\chi_q$.}
\label{TABLE0}
\end{table}

\begin{figure}[t]
\begin{tabular}{cc}
\includegraphics[width=8.5cm]{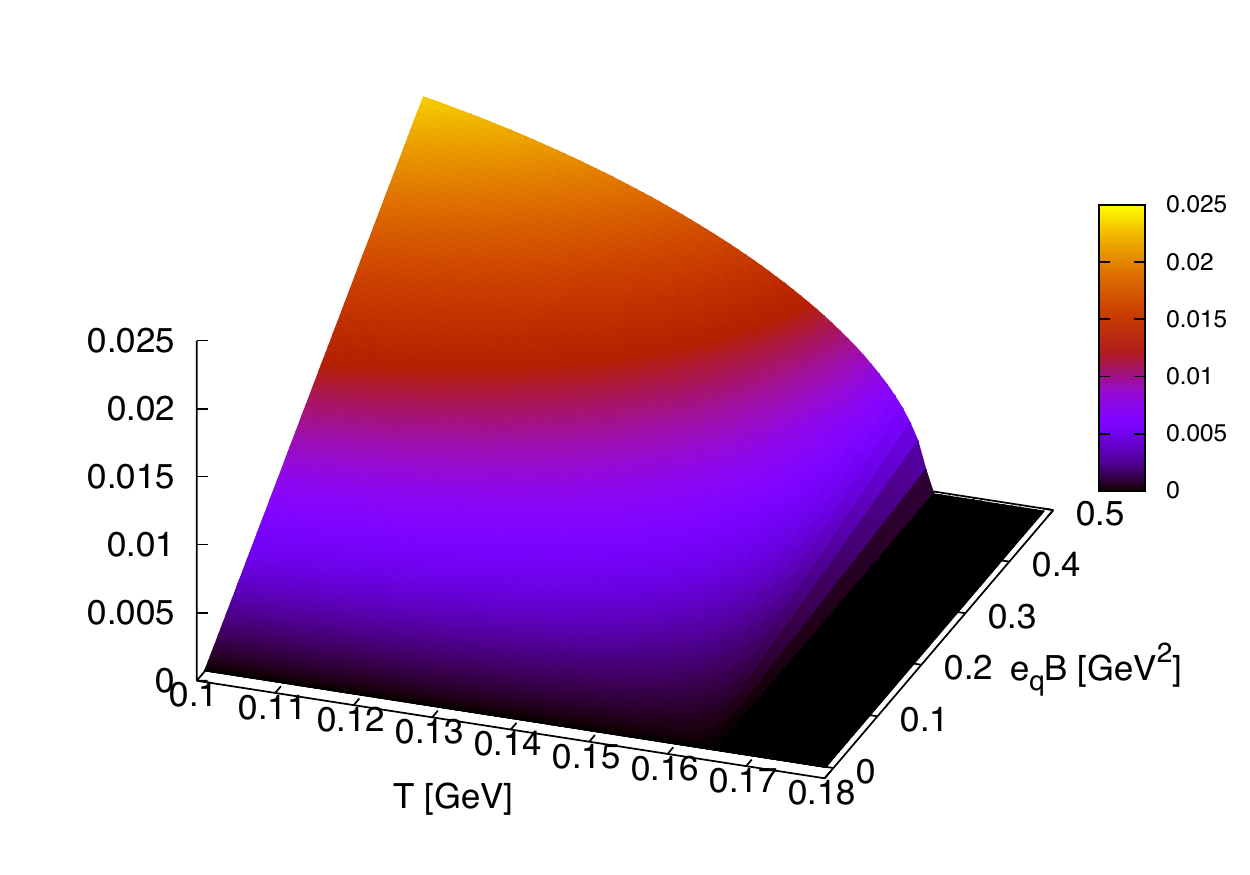}
\includegraphics[width=8.5cm]{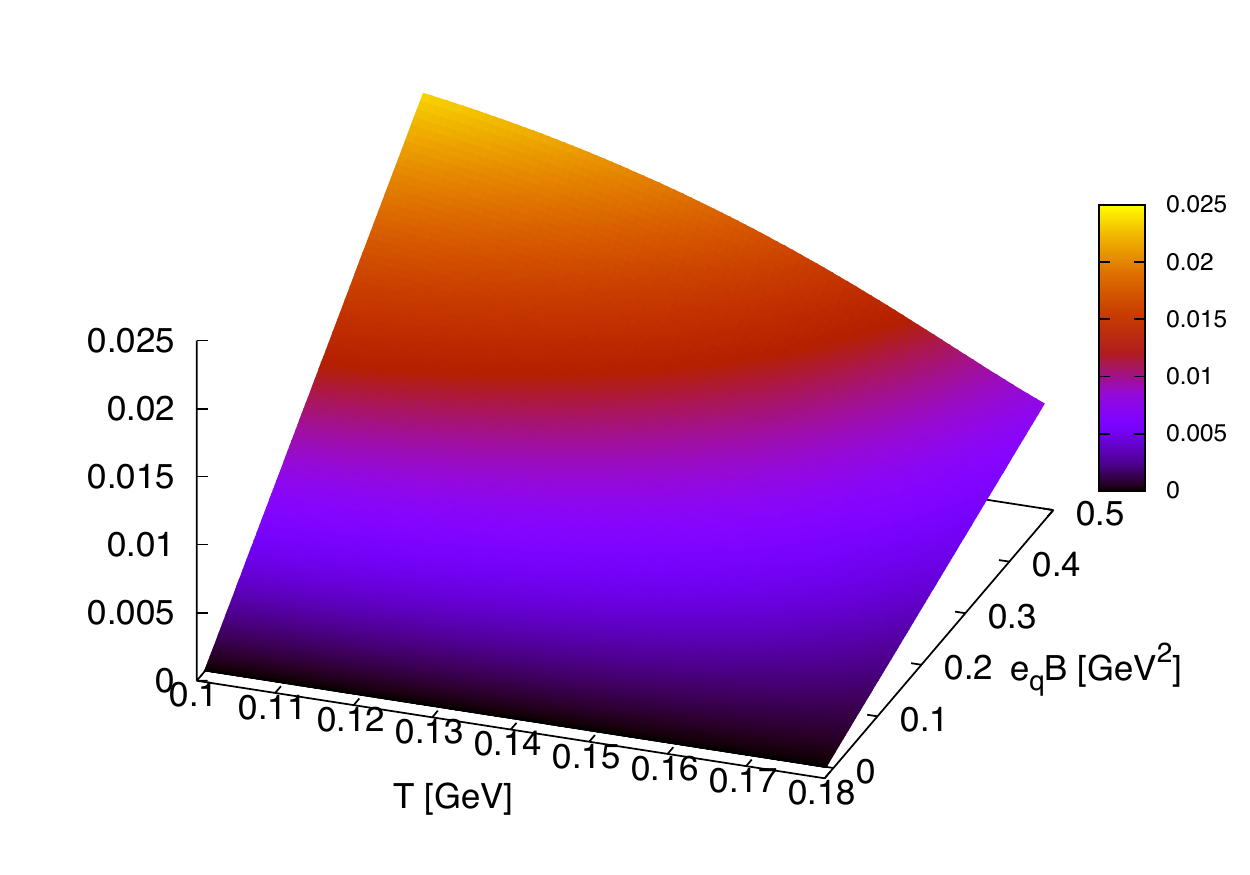}
\end{tabular}
\caption{(Color online) Tensor-polarization VEV (TP-VEV), $\langle q^\dagger\sigma_{ab}q\rangle_\mathrm{EM}$ [$\mathrm{GeV}^3$] as a function of $T$ and $e_qB$ for the chiral limit (left) and $m=5$ MeV (right). See the text for details.}       
\label{FIG67}
\end{figure}

Finally, we want to examine the external magnetic field dependence for TP-VEV in Eq.~(\ref{eq:VEV}). As already mentioned, in the leading order of $e_q$, TP-VEV is a linear function of the field strength tensor $F_{\mu\nu}$. Choosing a certain configuration for $F_{\mu\nu}$, we can write the following equation from Eq.~(\ref{eq:VEV}):
\begin{equation}
\label{eq:VEV10}
\langle q^{\dagger}\sigma_{ab}q\rangle_\mathrm{EM}
=e_qB\langle{q}^{\dagger}\Sigma q\rangle,
\end{equation}
where the Lorentz indices $a$ and $b$ are understood to pick up the magnetic field from the field strength tensor in Euclidean space. $B$ denotes the strength of the external magnetic field $B=|\bm{B}|$. Considering the positive quark charge, $u$ quark for instance, one can parameterize $e_qB$ as a real positive variable in the unit of $\mathrm{GeV}^2$. Note that the tensor condensate in the right-hand-side of Eq.~(\ref{eq:VEV10}) has been already computed as above. In the panel (b) of Figure~\ref{FIG45}, we show TP-VEV as functions of $e_qB$  for different $T$. The thick and thin lines denote the those for $m=0$ and $m=5$ MeV, respectively.  It turns out that, as $T$ increases, the slope of the lines decreases for the both current-quark masses. This behavior signals the (partial) chiral restoration, due to the decreasing of the tensor condensate, which plays the role of the slope of the line here. Note that this observation is consistent with that from the SU($3_c$) LQCD simulation~\cite{Bali:2012jv}, although the strength of the lines are different by about two times and the LQCD data show nonlinearity as $T$ increases which is not shown for the present leading-order calculations. The strength difference can be understood by the different renormalization scales, i.e. $\mu\approx0.85$ for ours and $\mu=2$ GeV for the LQCD simulation. If we go beyond the leading order $\mathrm{O}(e_q)$, it is sure that there appears nonlinearity in TP-VEV. However, we will not discuss this issue in the present work, and leave it for the future works. Comparing the lines for the chiral limit and finite quark mass, the line slope get diminished much for $m=0$ with respect to $T$, since TP-VEV also plays the role of the chiral order parameter. In other words, TP-VEV in the chiral limit becomes zero for even finite $e_qB$ at the chiral transition $T$. To see this situation clearly, in Figure~\ref{FIG67}, we show TP-VEV as functions of $eB$ as well as $T$ for $m=0$ (left) and $m=5$ MeV (right). As shown in the left panel, TP-VEV behaves linearly with respect to $eB$, and decreases its strength as $T$ increases. Then, it vanishes at $T_0=166$ MeV. In contrast, due to the crossover chiral phase transition, TP-VEV for $m=5$ MeV remains finite even beyond $T_0=170$ MeV. 

\section{Summary and conclusion}
In the present work, we have investigated the QCD magnetic susceptibility for SU($2_f$) as a function of temperature, beyond the chiral limit. For this purpose, we employed the liquid-instanton model (LIM), being modified by the trivial-holonomy caloron solution. We calculated the chiral $\langle iq^\dagger q\rangle$ and magnetic $\langle q^\dagger \Sigma q\rangle$ condensates,  QCD magnetic susceptibility $\chi_q$, and tensor-polarization VEV (TP-VEV) $\langle q^\dagger \sigma_{\mu\nu}\Sigma q\rangle_\mathrm{EM}$, numerically in Euclidean space. We compared our results with various theoretical estimations for those nonperturbative quantities. Important observations in the present work are given as follows:
\begin{itemize}
\item The chiral and tensor condensates manifest the role of the chiral order parameter, showing correct chiral restoration patters. We observe $\langle q^\dagger\Sigma q\rangle=(52,49)$ for $m=(0,5)$ MeV at $T=0$. These values are well comparable with the widely accepted one $\sim50$ MeV. The $T$ dependence of the tensor condensates are also compared with the LQCD data, and show qualitative agreement. At the chiral phase transition $T$, $T_0=170$ MeV, the tensor condensate for $m\ne0$ becomes $21$ MeV, whereas it is zero for the chiral limit.  
\item We find that the magnetic susceptibility is a smoothly decreasing function of $T$. The typical values for them are estimated as $\chi_q=(3.03,2.77)\,\mathrm{GeV}^{-2}$ for $m=(0,5)$ MeV at $T=0$. Again, these values are well compatible with the results from the LQCD and other effective models. Beyond $T_0=170$ MeV, the curve for the magnetic susceptibility behave almost as a linearly decreasing one with respect to $T$. At $T_0$, we observe about $20\%$ decreases in their strengths, in comparison to those at $T=0$.
\item TP-VEV is a linearly increasing function of the external magnetic field ($e_qB$) in the leading order. The slope of the TP-VEV line with respect to $e_qB$ decreases as $T$ increases, since the tensor condensate plays the role for its slope, signaling the (partial) restoration of SB$\chi$S. In that way, TP-VEV can be considered as a chiral order parameter. Hence, we find that it vanishes at $T_0=166$ MeV in the chiral limit, whereas it remains finite beyond $T_0$ for $m\ne0$, due to the crossover chiral phase transition.
\end{itemize}

From the above observations, we can conclude that the present model calculations have revealed meaningful and reliable results. Since the effects of the EM fields to QCD vacuum has been one of the most energetically progressing objects nowadays, it is meaningful to study more various nonperturbative quantities, being sensitive to the gauge field, such as the mixed quark-gluon condensate $\langle\bar{q}\sigma\cdot G q\rangle_\mathrm{EM}$, where $G_{\mu\nu}$ represents the gluon field strength tensor, for instance. Moreover, as discussed in the previous Section, calculations beyond the linearity on the magnetic field, i.e. beyond the leading order, can provide interesting modifications to the present results. Related works are under progress and appear elsewhere.  
\section*{Acknowledgments}
The author sincerely appreciates  that G.~Endr\"odi (Regensburg) kindly provided the lattice simulation data of Ref.~\cite{Bali:2012jv}. He also thanks C.~W.~Kao (CYCU), M.~M.~Musakhanov (Uzbekistan), and H.~-Ch.~Kim (Inha) for fruitful discussions. The numerical calculations were partially performed via the computing server ABACUS2 at KIAS.   
\section*{Appendix}



\begin{thebibliography}{99}
\bibitem{Abelev:2008ab}
  B.~I.~Abelev {\it et al.}  [STAR Collaboration],
  Phys.\ Rev.\ C {\bf 79}, 034909 (2009).
\bibitem{Bzdak:2011yy}
  A.~Bzdak and V.~Skokov,
  Phys.\ Lett.\ B {\bf 710}, 171 (2012).
\bibitem{Tuchin:2013ie}
  K.~Tuchin,
  arXiv:1301.0099 [hep-ph].
\bibitem{Fukushima:2008xe}
  K.~Fukushima, D.~E.~Kharzeev and H.~J.~Warringa,
  Phys.\ Rev.\  D {\bf 78}, 074033 (2008).
\bibitem{Kim:2004hd} 
  H.~-Ch.~Kim, M.~Musakhanov and M.~Siddikov,
  Phys.\ Lett.\ B {\bf 608}, 95 (2005).
\bibitem{Ioffe:1983ju} 
  B.~L.~Ioffe and A.~V.~Smilga,
  Nucl.\ Phys.\ B {\bf 232}, 109 (1984).
\bibitem{Ball:2002ps}
  P.~Ball, V.~M.~Braun and N.~Kivel,
  Nucl.\ Phys.\  B {\bf 649}, 263 (2003).
\bibitem{Bali:2013esa} 
  G.~S.~Bali, F.~Bruckmann, G.~Endrodi, F.~Gruber and A.~Schaefer,
  arXiv:1303.1328 [hep-lat].
\bibitem{Braun:2002en}
  V.~M.~Braun, S.~Gottwald, D.~Y.~Ivanov, A.~Schafer and L.~Szymanowski,
  Phys.\ Rev.\ Lett.\  {\bf 89}, 172001 (2002).
\bibitem{Belyaev:1984ic}
  V.~M.~Belyaev and Y.~I.~Kogan,
  Yad.\ Fiz.\  {\bf 40}, 1035 (1984).
\bibitem{Balitsky:1985aq}
  I.~I.~Balitsky, A.~V.~Kolesnichenko and A.~V.~Yung,
  Sov.\ J.\ Nucl.\ Phys.\  {\bf 41}, 178 (1985)
  [Yad.\ Fiz.\  {\bf 41}, 282 (1985)].
\bibitem{Frasca:2011zn} 
  M.~Frasca and M.~Ruggieri,
  Phys.\ Rev.\ D {\bf 83}, 094024 (2011).
\bibitem{Gorsky:2009ma} 
  A.~Gorsky and A.~Krikun ,
  Phys.\ Rev.\ D {\bf 79}, 086015 (2009).
\bibitem{Bergman:2008sg} 
  O.~Bergman, G.~Lifschytz and M.~Lippert,
  JHEP {\bf 0805}, 007 (2008).
\bibitem{Vainshtein:2002nv} 
  A.~Vainshtein,
  Phys.\ Lett.\ B {\bf 569}, 187 (2003).
\bibitem{Buividovich:2009ih} 
  P.~V.~Buividovich, M.~N.~Chernodub, E.~V.~Luschevskaya and M.~I.~Polikarpov,
  Nucl.\ Phys.\ B {\bf 826}, 313 (2010).
\bibitem{Braguta:2010ej} 
  V.~V.~Braguta {\it et al.},
  Phys.\ Atom.\ Nucl.\  {\bf 75}, 488 (2012).
\bibitem{Bali:2012jv} 
  G.~S.~Bali {\it et al.}, 
  Phys.\ Rev.\ D {\bf 86}, 094512 (2012).
\bibitem{Diakonov:2002fq}
  D.~Diakonov,
  Prog.\ Part.\ Nucl.\ Phys.\  {\bf 51}, 173 (2003).
\bibitem{Schafer:1995pz}
  T.~Schafer and E.~V.~Shuryak,
  Phys.\ Rev.\  D {\bf 53}, 6522 (1996).
\bibitem{Nam:2008ff} 
  S.~i.~Nam, H.~Y.~Ryu, M.~M.~Musakhanov and H.~-Ch.~Kim,
  J.\ Korean Phys.\ Soc.\  {\bf 55}, 429 (2009).
\bibitem{Nam:2011vn}
  S.~i.~Nam and C.~W.~Kao,
  Phys.\ Rev.\ D {\bf 83}, 096009 (2011).
\bibitem{Nam:2013fpa} 
  S.~i.~Nam and C.~W.~Kao,
  arXiv:1304.0287 [hep-ph].
\bibitem{Harrington:1976dj}
  B.~J.~Harrington and H.~K.~Shepard,
  Nucl.\ Phys.\  B {\bf 124}, 409 (1977).
\bibitem{Diakonov:1988my}
  D.~Diakonov and A.~D.~Mirlin,
  Phys.\ Lett.\  B {\bf 203}, 299 (1988).
\bibitem{Nam:2009nn}
  S.~i.~Nam,
  J.\ Phys.\ G {\bf 37}, 075002 (2010).
\bibitem{Beringer:1900zz} 
  J.~Beringer {\it {\it et al.}}  [Particle Data Group Collaboration],
  Phys.\ Rev.\ D {\bf 86}, 010001 (2012).
\bibitem{Schwinger:1951nm}
  J.~S.~Schwinger,
  Phys.\ Rev.\  {\bf 82}, 664 (1951).
\bibitem{Schafer:1996wv}
  T.~Schafer and E.~V.~Shuryak,
  Rev.\ Mod.\ Phys.\  {\bf 70}, 323 (1998).
\bibitem{Goeke:2007bj}
  K.~Goeke, M.~M.~Musakhanov and M.~Siddikov,
  Phys.\ Rev.\ D {\bf 76}, 076007 (2007)
\bibitem{sinam}
S.~i.~Nam, in preparation.
\bibitem{Goeke:2007nc} 
  K.~Goeke, H.~-Ch.~Kim, M.~M.~Musakhanov and M.~Siddikov,
  Phys.\ Rev.\ D {\bf 76}, 116007 (2007).
\end{thebibliography}
\end{document}